\documentclass{article}
\usepackage[T1]{fontenc} 
\usepackage[utf8]{inputenc} 
\usepackage{amsmath,cite,url}
\usepackage{graphicx}
\usepackage[lbd]{ismir}
\usepackage{kotex}
\usepackage{color}

\usepackage[firstpage]{draftwatermark}
\definecolor{lightgray}{rgb}{0.9,0.9,0.9}
\definecolor{darkgray}{rgb}{0.4,0.4,0.4}
\SetWatermarkFontSize{12pt}
\SetWatermarkScale{1.1}
\SetWatermarkAngle{90}
\SetWatermarkHorCenter{202mm}
\SetWatermarkVerCenter{170mm}
\SetWatermarkColor{darkgray}
\SetWatermarkText{Late-Breaking / Demo Session Extended Abstract, ISMIR 2025 Conference}

\title{Motive-level Analysis of Form-functions Association \\in Korean Folk song}

\threeauthors
  {Danbinaerin Han} {KAIST \\ {\tt naerin71@kaist.ac.kr}}
  {Dasaem Jeong} {Sogang University \\ {\tt dasaemj@sogang.ac.kr}}
  {Juhan Nam} {KAIST \\ {\tt juhan.nam@kaist.ac.kr}}

\def\authorname{D. Han, D. Jeong, and J. Nam}

\begin{document}

\maketitle
\begin{abstract}

Computational analysis of folk song audio is challenging due to structural irregularities and the need for manual annotation. We propose a method for automatic motive segmentation in Korean folk songs by fine-tuning a speech transcription model on 
audio lyric with motif boundary annotation. 
Applying this to 856 songs, we extracted motif count and duration entropy as structural features. Statistical analysis revealed that these features vary systematically according to the social function of the songs.
Songs associated with collective labor, for instance, showed different structural patterns from those for entertainment or personal settings. This work offers a scalable approach for quantitative structural analysis of oral music traditions.

\end{abstract}

\section{Introduction}\label{sec:introduction}

Folk songs emerge from the contexts of daily life and have been passed down through generations. 
This origin suggests the possibility of a strong correlation between the function of folk songs and their musical forms.
The early computational ethnomusicologist Alan Lomax supported the claim that the functional/social context determines the form of a song by attempting to quantify the musical characteristics of thousands of folk songs and statistically link them to cultural contexts\cite{lomax1962song}. 
Subsequent studies have focused on defining musical features that capture the structural aspects of folk songs or have employed more recent statistical methods to explore the relationship between the context of a song's appearance and its various musical features\cite{savage2012cantocore, savage2015statistical, mehr2019universality}.

A key challenge in computational analyzing folk music lies in the nature of its audio recordings. Folk songs are typically sourced from field recordings featuring non-professional singers, resulting in music with considerable variability in pitch, timing, and formal structure. 
Although motif segmentation for structural analysis is a traditional task in the MIR field, segmentation principles\cite{nieto2020audio}, which are based on western music theory, are not well-suited for monophonic, non-Western music. 
While a method for automatically segmenting motifs in traditional folk songs by considering melodic breaks and segment length regularity has been proposed\cite{kroher2017discovery}, it is difficult to apply to song samples with highly irregular rhythms.


In this study, we propose an approach that automatically segments motives based on lyrics by fine-tuning a automatic speech recognition model on folk song audio. 
Then, we apply this model to analyze structural properties of Korean folk song and the relationship between the function of songs and their musical characteristics.




\section{Methodology}
\begin{figure}[!t]               
  \centering                       
  \includegraphics[width=0.91\columnwidth]
    {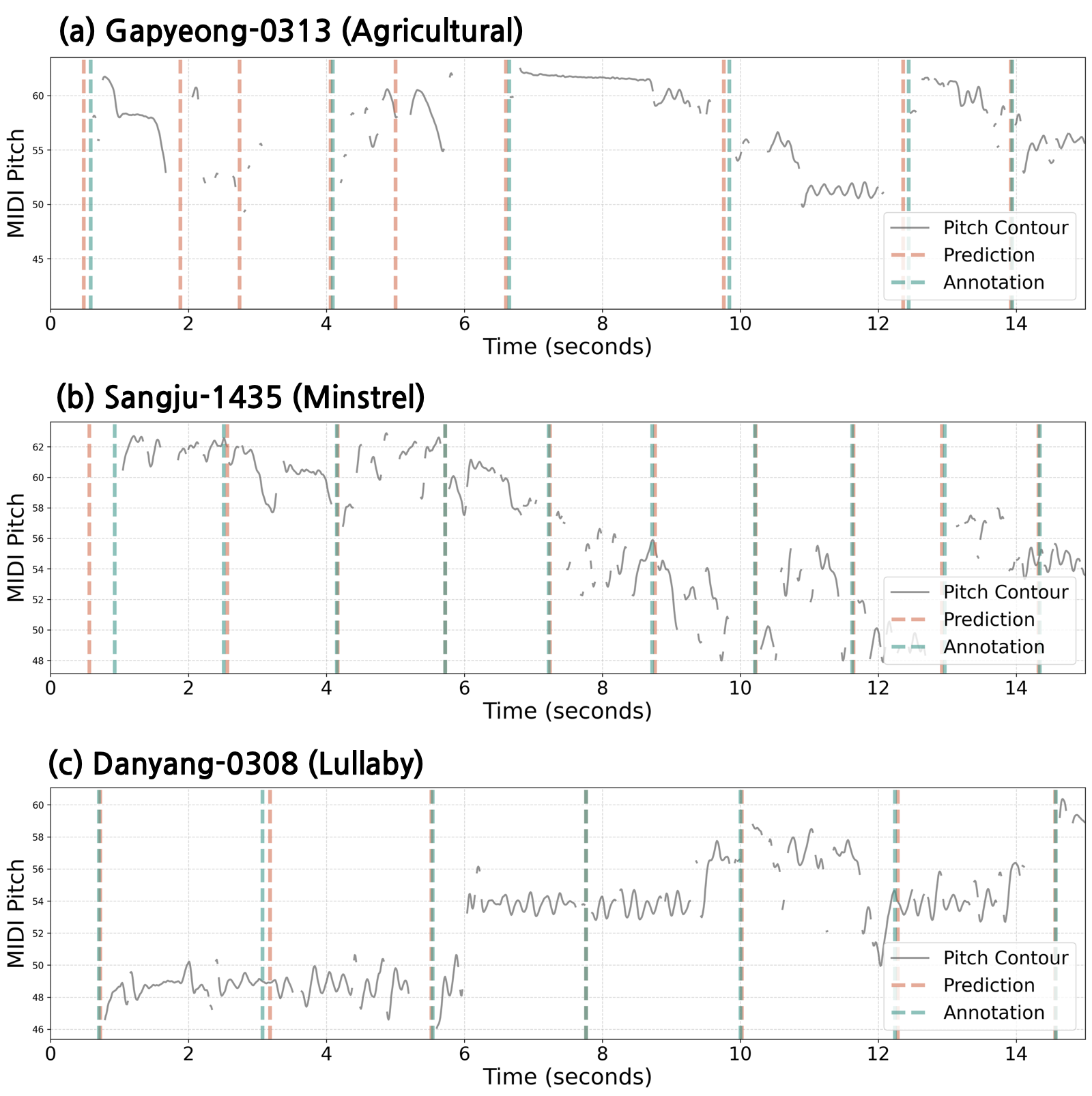}
  \caption{Comparison of model-predicted and human-annotated motif boundaries on the pitch contours of three folk song examples: (a) agricultural, (b) minstrel, and (c) lullaby.}
  \label{fig:phrases}     
\end{figure}

\subsection{Korean Folk Song Dataset}
This study is based on a dataset of 856 Korean folk songs, selected from a curated collection of 2,255 tracks in the ``Anthology of Korean Traditional Folksongs''—a comprehensive archive compiled by Munhwa Broadcasting Corporation (MBC). The audio files and metadata were obtained from a public website via web crawling (see ~\cite{urisori, han2023finding}). We retained only monophonic vocal tracks longer than 30 seconds for analysis. 



\subsection{Manual Annotation with Transcribed Lyrics}
Based on the empirical observation of a strong correlation between lyrical chunks and musical motives in folk songs, we found that segmenting motives based on lyrical content—considering both semantics and syllable count—provides more reliable results than segmentation based on audio features alone.

A subset of approximately 150 songs was randomly selected from the main collection, prioritizing those of suitable length and quality,
with 10–20 songs chosen from each functional category. For these songs, motif boundaries were manually annotated by a single, trained annotator, with at least one minute of audio processed for each track while referring to the transcribed lyrics. This resulted in a dataset of 186.4 minutes of annotated audio, encompassing a variety of song forms.
Finally, the original lyric transcriptions were manually corrected and aligned with the audio and boundary annotations.
%

 

\subsection{Whisper Fine Tuning}

To automatically detect the boundaries of musical phrases, we fine-tuned the Whisper speech recognition model \cite{radford2023robust}. This approach is based on the hypothesis that the capabilities of an ASR model to understand semantic and prosodic information are analogous to the process of identifying musical phrase boundaries.
We specifically fine-tuned the whisper-small model on our manually annotated dataset, where phrase boundary timings were represented as special timestamp tokens.
 This training was conducted on two A5000 GPUs, employing a learning rate of 1e-5 and a batch size of 8.
 The model that yielded the lowest WER on the validation set at 100 iterations was selected for motif segmentation across the entire dataset.


\subsection{Motif Duration Analysis}

To analyze the structural characteristics of each song, we calculated two features for every one-minute segment of audio: 1) the motif count, representing the density of phrases, and 2) the entropy of motif durations, measuring their variability.
High entropy in motif duration signifies a wide range of motive lengths, suggesting an irregular or unpredictable musical flow. Conversely, low entropy indicates that motive durations are more uniform or follow a recurring pattern, which is often associated with a regular meter or a highly structured composition.
To compute the entropy, the duration of each motive was first log-transformed (base 2). We then created a histogram of these values with a bin width of 0.2 and calculated the entropy from the resulting distribution to quantify the variability of motive lengths. This process quantified the distribution of motive lengths within each one-minute interval.


\section{Results}

\begin{figure}[!t]               
  \centering                       
  \includegraphics[width=\columnwidth]
    {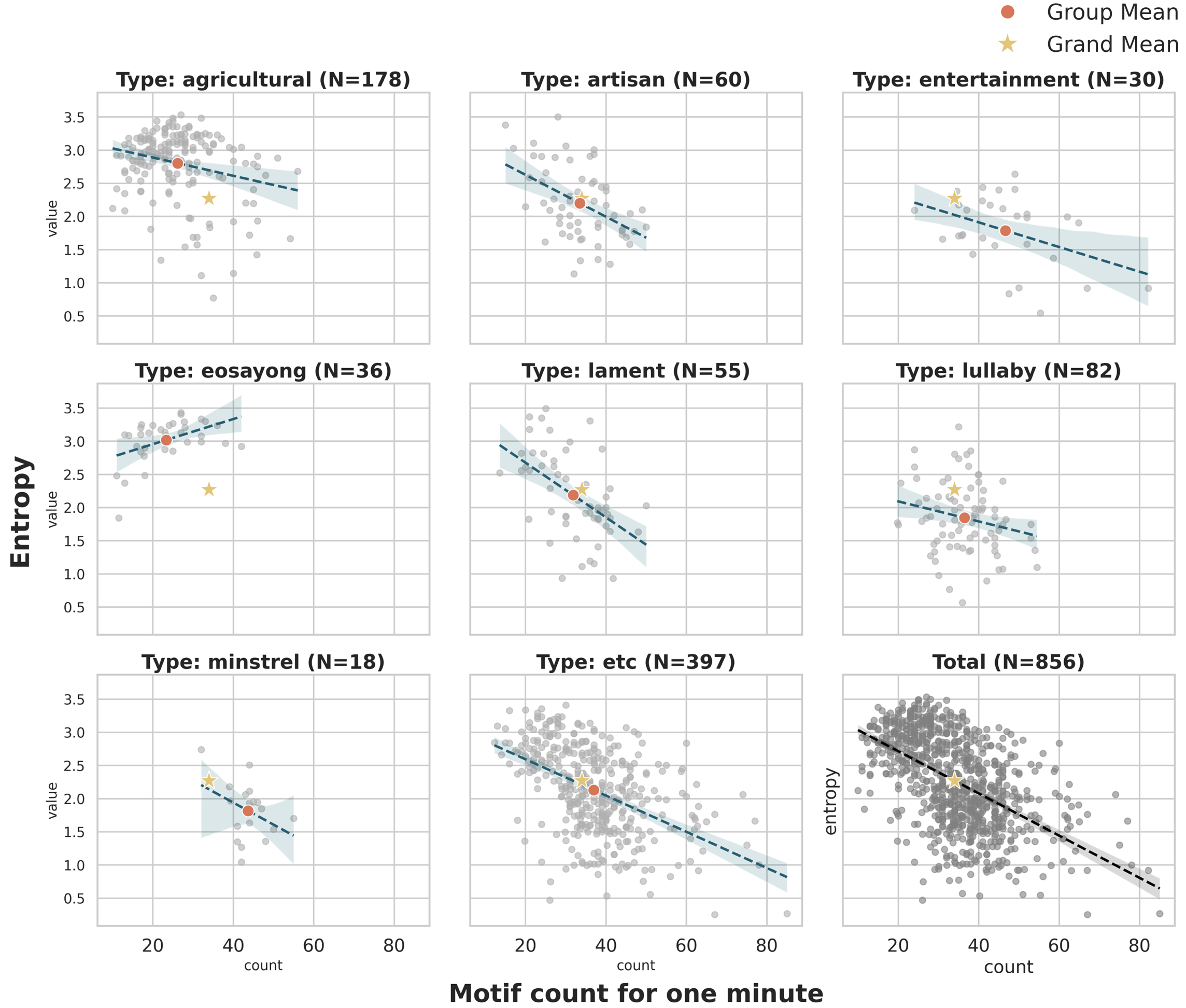}
  \caption{Scatter plots showing the relationship between motif count per minute and the entropy of motif durations for each of the seven functional categories. }
  \label{fig:scatters}     
\end{figure}

We evaluated the automatic segmentation model against our manually annotated dataset via 10-fold cross-validation. The model achieved a mean F1-score of 0.820 for boundary detection (0.1s tolerance), with performance varying by genre from 0.710 (agricultural) to 0.880 (artisan songs). Figure \ref{fig:phrases} compares model-predicted motif boundaries with human annotations over the pitch contours of three folk-song examples.


Next, we applied the trained model to the entire dataset of 856 songs to automatically segment them. We then conducted a Multivariate Analysis of Variance to examine the relationship between song function and our two structural features: motif count and duration entropy. The analysis revealed a statistically significant effect of song function on the combined dependent variables (Wilks' Λ = 0.68, F(14, 1532) = 23.52, p < 0.001). This result indicates that the functional purpose of a folk song is strongly associated with the density and variability of its musical motives as shown in Figure~\ref{fig:scatters}.

Post-hoc tests revealed distinct structural patterns across different functional categories of folk songs.
\textbf{Agricultural} songs (work songs for farming, Fig. \ref{fig:phrases}a) are characterized by a high motif count and high entropy. This profile differs from that of \textbf{artisan} songs (work songs performed indoors), which exhibit average counts and entropy, likely reflecting the different social and physical dynamics of the labor contexts.
A particularly interesting contrast was found in songs of lament. \textbf{Eosayong} (woodcutters' laments, typically sung by men) displayed the lowest motif count but high entropy. This stands in sharp contrast to \textbf{lament} songs performed by women, which showed different patterns.
In contrast, \textbf{entertainment} and \textbf{minstrel} songs exhibit a high motif count and low entropy (Fig. \ref{fig:phrases}b). This suggests the use of clear and regular rhythmic structures, which are effective for engaging an audience or pacing a collective activity.
Finally, \textbf{lullabies} show a moderate motif count and low entropy (Fig. \ref{fig:phrases}c). This combination points to a preference for stable and predictable melodic structures over complex ones, which is well-suited for the personal and tranquil context of soothing a child to sleep.
\bibliography{ISMIRtemplate}

%
%
%
%
%

\end{document}